\documentclass[12pt]{article}
\usepackage{epsfig}
\usepackage{subfigure}
\usepackage{amsmath}
\usepackage{amsfonts}
\usepackage{amssymb}
\usepackage{url}
\usepackage{hyperref}
\newcommand{\be}{\begin{equation}}
\newcommand{\ee}{\end{equation}}
\newcommand{\bea}{\begin{eqnarray}}
\newcommand{\eea}{\end{eqnarray}}

\begin{document}
\begin{titlepage}
\begin{flushright}
\end{flushright}
\vspace{4\baselineskip}
\begin{center}
{\Large\bf 
A simple 5D SO(10) GUT and sparticle masses  
}
\end{center}
\vspace{1cm}
\begin{center}
{\large Takeshi Fukuyama $^{a,b,}$
\footnote{\tt E-mail:fukuyama@se.ritsumei.ac.jp} 
and Nobuchika Okada $^{c,d,}$
\footnote{\tt E-mail:okadan@post.kek.jp}}
\end{center}
\vspace{0.2cm}
\begin{center}
${}^{a}$ {\small \it Department of Physics and R-GIRO, 
Ritsumeikan University, Kusatsu, Shiga, 525-8577, Japan}\\
${}^{b}$ {\small \it SISSA, Trieste I-34014, Italy}\\
${}^{c}$ {\small \it Department of Physics, 
University of Maryland, College Park, MD 20742, USA}\\ 
${}^{d}$ {\small \it Theory Group, KEK,
Oho 1-1, Tsukuba, 305-0801, Japan}\\

\medskip
\vskip 10mm
\end{center}
\vskip 10mm
\begin{abstract}
Simple supersymmetric SO(10) GUT in five dimensions is proposed, 
 in which the fifth dimension is compactified 
 on the $S^1/(Z_2\times Z_2^\prime)$ orbifold 
 with two inequivalent branes at the orbifold fixed points. 
In this model, all matter and Higgs multiplets reside 
 on one brane (PS brane) where the Pati-Salam (PS) symmetry is manifest, 
 while only the SO(10) gauge multiplet resides on the bulk. 
The supersymmetry breaking on the other brane (SO(10) brane) is 
 transmitted to the PS brane through the gaugino mediation 
 with the bulk gauge multiplet. 
We examine sparticle mass spectrum in this setup and 
 show that the neutralino LSP as the dark matter candidate 
 can be realized when the compactification scale of 
 the fifth dimension is higher than the PS symmetry breaking scale, 
 keeping the successful gauge coupling unification 
 after incorporating threshold corrections 
 of Kaluza-Klein modes of the bulk gauge multiplets. 

\end{abstract}
\end{titlepage}

\setcounter{page}{1}
\setcounter{footnote}{0}

\section{Introduction}

Supersymmetric (SUSY) SO(10) Grand Unified Theory (GUT) 
 is one of the most promising candidates beyond the Standard Model (SM).
Among several models, 
 the so-called renormalizable minimal SO(10) model has been paid 
 a particular attention, where two kinds of Higgs multiplets 
 $\{{\bf 10} \oplus {\bf \overline{126}}\}$ 
 are utilized for the Yukawa couplings with matters 
 ${\bf 16}_i~(i=\mbox{generation})$ 
 \cite{Babu:1992ia, M-K-F, Bajc:2002iw}.
A remarkable feature of the model is its high predictivity. 
Indeed, it fixes all quark-lepton mass matrices including 
 heavy right-handed neutrinos. 
So, it predicts a wide range of phenomena in low energy physics 
 as well as new phenomena beyond SM, such as 
 lepton-flavor violations, leptogenesis, proton decay etc. 
The recent reviews of the applications of minimal SO(10) 
 to these wide range are given in \cite{F-O-01}. 
However, after KamLAND data \cite{Eguchi:2002dm} was released, 
 it entered to the precision measurements, 
 and many authors performed new data fitting 
 to match up these new data \cite{Goh:2003hf, Bertolini:2006pe}.

On the other hand, there has been another theoretically 
 important progress in the Higgs sector of the SO(10) model.
The symmetry breaking pattern of the simple renormalizable 
Higgs superpotential \cite{clark} down to the SM was analyzed in detail
 \cite{Fukuyama:2004xs, Bajc:2004xe}. 
This gives the very unambiguous and detailed structures 
 between the GUT and the Standard Model in unprecedented ways. 
However it accommodates conflicts 
 on the gauge coupling unifications etc. \cite{Bertolini:2006pe}.

In addition to the issue of the gauge coupling unification and proton decay, 
 the minimal SO(10) model potentially suffers from the problem 
 that the gauge coupling blows up around the GUT scale. 
This is because the model includes many Higgs multiplets of 
 higher dimensional representations. 
In field theoretical point of view, this fact implies 
 that the GUT scale is a cutoff scale of the model, 
 and more fundamental description of the minimal SO(10) model 
 would exist above the GUT scale. 

Not only to solve these problems but also to give solutions 
 to hitherto unsolved problems like the origin of SUSY breaking 
 mediations, we have considered GUT in five dimensions (5D).  
As a simple realization of such a scenario, 
 the minimal SO(10) model was considered in warped extra dimensions \cite{F-K-O}. 
In this scenario, the Anti-de Sitter curvature and the fifth dimensional 
 radius were chosen so as to realize the GUT scale 
 as an effective cutoff scale in 4-dimensional effective theory. 
This idea has been utilized in an extended model proposed in \cite{MOY}, 
 where the so-called type II seesaw mechanism dominates 
 to realize the tiny neutrino masses through the warped geometry. 
In both models, the SO(10) gauge symmetry was considered 
 to be broken by Higgs multiplets on branes 
 as usual in 4-dimensional models.

Another possibility of constructing GUT models in extra-dimensions 
 is to consider the so-called orbifold GUT \cite{Kawamura},  
 where the GUT gauge symmetry is (partly) broken by 
 orbifold boundary conditions.  
A class of SO(10) models in 5D has been proposed \cite{Raby}, 
 where SO(10) gauge symmetry is broken into the PS
 gauge group by orbifold boundary condition on $S^1/(Z_2\times Z_2')$ 
 and further symmetry breaking into the SM gauge group 
 is achieved by VEVs of Higgs multiplets on a brane. 
In this context, we have recently proposed 
 a very simple SO(10) model \cite{F-O1}. 
In this model, all matter and Higgs multiplets reside only 
 on a brane (PS brane) where the PS gauge symmetry is manifest, 
 so that low energy effective description of this model 
 is nothing but the PS model in 4D with a special set of 
 matter and Higgs multiplets. 
At energies higher than the compactification scale ($M_c$), 
 the Kaluza-Klein modes of the bulk SO(10) gauge multiplet 
 are involved in the particle contents and in fact, 
 the gauge coupling unification was shown to be successfully realized 
 by incorporating the Kaluza-Klein mode threshold corrections 
 into the gauge coupling running \cite{F-O1}. 
More recently, it has been shown \cite{F-O-O} that 
 this orbifold GUT model is applicable to 
 the smooth hybrid inflation \cite{SmoothHI}.  
Interestingly, the inflation model can fit 
 the WMAP 5-year data \cite{WMAP} very well 
 by utilizing the PS breaking scale ($v_{PS}$) and 
 the gauge coupling unification scale predicted 
 independently of cosmological considerations.

In any SUSY models, the origin of SUSY breaking and 
 its mediation to the minimal SUSY SM (MSSM) sector 
 is an important issue. 
Since the flavor-dependent soft SUSY breaking terms 
 are severely constrained by the current experiments, 
 a mechanism to transmit SUSY breaking naturally 
 in a flavor-blind way is favorable . 
From this motivation, the SUSY breaking sector on 
 the other brane is examined in Ref.~\cite{F-O1}. 
In this setup, the bulk gauginos first obtain masses, 
 and sfermion masses on the PS brane are automatically 
 generated through the renormalization group equation 
 (RGE) from the compactification scale to the electroweak scale. 
Importantly, the sfermion masses generated in this way 
 are flavor-blind, because the interaction transmitting 
 the gaugino mass to sfermion masses is the gauge interaction. 
This scenario is nothing but the gaugino mediation \cite{gMSB}.

Unfortunately, the simple setup taken in Ref.~\cite{F-O1} 
 results in the right-handed slepton (normally, stau) 
 being the lightest superpartner (LSP) 
 and is disfavored in cosmological point of view. 
In this paper, we show that the previous conclusion 
 is the result from the special condition, $v_{PS}=M_c$, 
 adopted in \cite{F-O1} and if we relax it to be $v_{PS} < M_c$, 
 bino-like neutralino arises as the LSP and so 
 a usual dark matter candidate in SUSY models.  
We also show that even for the parameter choice, $v_{PS} < M_c$, 
 the gauge coupling unification is still successfully realized.

The paper is organized as follows: 
In the next section we briefly review the setup of 
 our orbifold SO(10) GUT model. 
In Sec.~3, we investigate the gauge coupling unification 
 for the case $v_{PS} < M_c$. 
In Sec.~4, we examine the gaugino mediation in our model  
 for $v_{PS} < M_c$ and investigate sparticle masses, 
 in particular, masses of the right-handed slepton (stau) and bino. 
We will see that bino becomes the LSP 
 when $M_c$ is sufficiently large. 
The last section is devoted for conclusion.

\section{Model setup}
Here we briefly review the orbifold SO(10) GUT model 
 proposed in Ref.~\cite{F-O1}. 
The model is described in 5D and the 5th dimension is compactified 
 on the orbifold $S^1/{Z_2 \times Z_2^\prime}$ \cite{Kawamura}. 
A circle $S^1$ with radius $R$ is divided by 
 a $Z_2$ orbifold transformation $y \to -y$ 
 ($y$ is the fifth dimensional coordinate $ 0 \leq y < 2 \pi R$)
 and this segment is further divided by a $Z_2^\prime$ transformation 
 $y^\prime \to -y^\prime $ with $y^\prime = y + \pi R/2$. 
There are two inequivalent orbifold fixed points at $y=0$ and $y=\pi R/2$. 
Under this orbifold compactification, a general bulk wave function 
 is classified with respect to its parities,  
 $P=\pm$ and $P^\prime=\pm$, under $Z_2$ and $Z_2^\prime$, respectively.

Assigning the parity ($P,P^\prime $) to 
 the bulk SO(10) gauge multiplet as listed in Table~1, 
 only the PS gauge multiplet has zero-mode and 
 the bulk 5D N=1 SUSY SO(10) gauge symmetry is broken 
 to 4D N=1 SUSY PS gauge symmetry. 
Since all vector multiplets has wave functions  
 on the brane at $y=0$, the SO(10) gauge symmetry is respected there, 
 while only the PS symmetry is on the brane at $y=\pi R/2$ (PS brane). 

\begin{table}[h]
\begin{center}
\begin{tabular}{|c|c|c|}
\hline
$(P,P')$ & bulk field & mass\\
\hline 
& & \\
$(+,+)$ & $V(15,1,1)$, $V(1,3,1)$, $V(1,1,3)$ & $\frac{2n}{R}$\\
& & \\
\hline
& & \\
$(+,-)$ &  $V(6,2,2)$ & $\frac{(2n+1)}{R}$ \\
& & \\
\hline
& & \\
$(-,+)$ &  $\Phi (6,2,2)$
& $\frac{(2n+1)}{R}$\\
& & \\
\hline
& & \\
$(-,-)$ & $\Phi (15,1,1)$, $\Phi (1,3,1)$, $\Phi (1,1,3)$ & $\frac{(2n+2)}{R}$ \\
& & \\
\hline
\end{tabular}
\end{center}
\caption{
 ($P,~P^\prime$) assignment and masses ($n \geq 0$) of fields 
 in the bulk SO(10) gauge multiplet $(V,~\Phi)$ 
 under the PS gauge group. 
$V$ and $\Phi$ are the vector multiplet and adjoint chiral 
 multiplet in terms of 4D N=1 SUSY theory. 
}
\label{t1}
\end{table}

We place all the matter and Higgs multiplets on the PS brane, 
 where only the PS symmetry is manifest 
 so that the particle contents are in the representation 
 under the PS gauge symmetry, not necessary to be 
 in SO(10) representation.   
For a different setup, see \cite{Raby}. 
The matter and Higgs in our model is listed in Table~2. 
For later conveniences, let us introduce the following notations: 
\bea
H_1&=&({\bf 1},{\bf 2},{\bf 2})_H, ~H_1^{\prime}=({\bf 1},{\bf 2},{\bf 2})'_H,
\nonumber \\
H_6&=&({\bf 6},{\bf 1},{\bf 1})_H, ~H_{15}=({\bf 15},{\bf 1},{\bf 1})_H,
\nonumber \\ 
H_L&=&({\bf 4},{\bf 2},{\bf 1})_H,
~\bar{H}_L =(\overline{{\bf 4}},{\bf 2},{\bf 1})_H,  
\nonumber \\
H_R&=&({\bf 4},{\bf 1},{\bf 2})_H,
~\bar{H}_R=(\overline{\bf 4},{\bf 1},{\bf 2})_H.
\eea

Superpotential relevant for fermion masses is given by%
\footnote{
For simplicity, we have introduced only minimal terms 
 necessary for reproducing observed fermion mass matrices. 
}
\bea
W_Y&=& Y_{1}^{ij} F_{Li} F_{R j}^c H_1 
+\frac{Y_{15}^{ij}}{M_5} F_{Li} F_{R j}^c 
 \left(H_1^{\prime} H_{15} \right) \nonumber\\ 
&+&\frac{Y_R^{ij}}{M_5} F_{Ri}^c F_{R j}^c 
 \left(H_R H_R \right) +\frac{Y_L^{ij}}{M_5} F_{Li}F_{L j} 
 \left(\bar{H}_L \bar{H}_L \right), 
\label{Yukawa}
\eea 
where $M_5$ is the 5D Planck scale. 
The product, $H_1^{\prime} H_{15}$, effectively works 
 as $({\bf 15},{\bf 2},{\bf 2})_H$, 
 while $H_R H_R$ and $\bar{H}_L \bar{H}_L$ 
 effectively work as $({\bf 10},{\bf 1},{\bf 3})$ and 
 $(\overline{{\bf 10}},{\bf 3},{\bf 1})$, respectively, 
 and are responsible for the left- and the right-handed 
 Majorana neutrino masses. 
Providing VEVs for appropriate Higgs multiplets, 
 fermion mass matrices are obtained. 
There are a sufficient number of free parameters 
 to fit all the observed fermion masses and mixing angles. 

\begin{table}[h]
{\begin{center}
\begin{tabular}{|c|c|}
\hline
& brane at $y=\pi R/2$ \\ 
\hline
& \\
Matter Multiplets & $\psi_i=F_{Li} \oplus F_{Ri}^c \quad (i=1,2,3)$ \\
 & \\
\hline
 & \\
Higgs Multiplets & 
$({\bf 1},{\bf 2},{\bf 2})_H$,  
$({\bf 1},{\bf 2},{\bf 2})'_H$,
$({\bf 15},{\bf 1},{\bf 1})_H$,
$({\bf 6},{\bf 1},{\bf 1})_H$ \\  & 
$({\bf 4},{\bf 1},{\bf 2})_H$, 
$(\overline{{\bf 4}},{\bf 1},{\bf 2})_H$, 
$({\bf 4},{\bf 2},{\bf 1})_H$, 
$(\overline{{\bf 4}},{\bf 2},{\bf 1})_H$  \\
& \\
\hline
\end{tabular}
\end{center}}
\caption{
Particle contents on the PS brane. 
$F_{Li}$ and $F_{Ri}^c$ are matter multiplets 
 of i-th generation in $({\bf 4}, {\bf 2}, {\bf 1})$ and 
 $(\overline{\bf 4}, {\bf 1}, {\bf 2})$ representations, respectively. 
}
\end{table}

We introduce Higgs superpotential invariant under the PS symmetry 
 such as \cite{F-O1}%
\footnote{
It is possible to consider a different superpotential 
 by introducing a singlet chiral superfield \cite{F-O-O}, 
 so that this model is applicable to 
 the smooth hybrid inflation scenario \cite{SmoothHI}. 
}
\bea
W &=& 
 \frac{m_1}{2} H_1^2 + \frac{m_1^\prime}{2} H_1^{\prime 2} 
 + m_{15}~{\rm tr}\left[H_{15}^2 \right] 
  +m_4 \left(\bar{H}_L H_L+\bar{H}_R H_R\right) \nonumber\\
&+& 
\left(H_L \bar{H}_R+ \bar{H}_L H_R \right) 
\left( \lambda_1 H_1 + \lambda_1^\prime H_1^\prime \right) 
+\lambda_{15} \left(\bar{H}_R H_R + \bar{H}_L H_L\right) 
H_{15} \nonumber\\
&+&
\lambda~{\rm tr}\left[H_{15}^3 \right]+
\lambda_6 
 \left( H_L^2+ \bar{H}_L^2 + H_R^2 + \bar{H}_R^2 \right) 
 H_6 .
\label{HiggsW}
\eea
Parameterizing 
 $ \langle H_{15} \rangle =\frac{v_{15}}{2 \sqrt{6}} 
 {\rm diag}(1,1,1,-3)$, 
 SUSY vacuum conditions from Eq.~(\ref{HiggsW}) and 
 the D-terms are satisfied by solutions,  
\bea
v_{15} =\frac{2 \sqrt{6}}{3 \lambda_{15}} m_4,~~~
\langle           H_R  \rangle = 
\langle \bar{H}_R \rangle = 
\sqrt{
 \frac{8 m_4}{3 \lambda_{15}^2} 
 \left( m_{15} -\frac{\lambda}{\lambda_{15}} m_4 \right) }
 \equiv v_{PS} 
\eea 
and others are zero, by which the PS gauge symmetry is broken 
 down to the SM gauge symmetry.  
We choose the parameters so as to be 
 $ v_{15} \simeq \langle H_R \rangle = \langle \bar{H}_R \rangle$.  
Note that the last term in Eq.~(\ref{HiggsW}) is necessary 
 to make all color triplets in $H_R$ and $\bar{H}_R$ heavy.

The Higgs doublet mass matrix is given by
\begin{equation}
\left(
       \begin{array}{ccc}
        H_1 & H_1^\prime & H_L \end{array}
\right)\left(
        \begin{array}{ccc}
        m_1 &  0         & \lambda_1        \langle H_R \rangle \\
        0   & m_1^\prime & \lambda_1^\prime \langle H_R \rangle \\ 
        \lambda_1 \langle \bar{H}_R \rangle &  
        \lambda_1^\prime \langle \bar{H}_R \rangle & m_4 
        \end{array}
\right)\left(
        \begin{array}{c}
        H_1\\
        H_1^\prime \\
        \bar{H}_L
        \end{array}
\right).
\end{equation} 
Requiring the tuning of parameters to satisfy 
\bea
\det M=m_1 m_1^\prime m_4 - 
 (m_1 \lambda_1^{\prime 2} + m_1^\prime \lambda_1^2) v_{PS}^2=0, 
\label{vev}
\eea
only one pair of Higgs doublets out of the above three pairs 
 becomes light and is identified as the MSSM Higgs doublets 
 while the others have mass of the PS symmetry breaking scale.

In Ref.~\cite{F-O1}, assuming $M_c=v_{PS}$ and imposing 
 the left-right symmetry, the gauge coupling unification 
 was examined. 
We relax this assumption and consider the case $v_{PS} < M_c$ 
 in the next section. 
However, note that $v_{PS}=1.19 \times 10^{16}$ GeV 
 is fixed as the same as in Ref.~\cite{F-O1} 
 since the PS scale is determined as the scale 
 where the SU(2)$_L$ and SU(2)$_R$ gauge couplings 
 coincide with each other. 
The PS scale in our model is very high relative to 
 other 5D orbifold SO(10) models \cite{Raby}. 
The high value of $v_{PS}$ is advantageous 
 for dangerous proton decay due to dimension six operators. 
 From Eq.~(\ref{Yukawa}), the right-handed neutrino mass scale 
 is given by $M_R \sim Y_R v_{PS}^2/M_5$. 
For $M_5 \sim 10^{17}$ GeV 
 (which can be estimated from the parameters 
 obtained in the next section), 
 the scale $M_R = {\cal O}(10^{14}~\mbox{GeV})$ 
 preferable for the seesaw mechanism can be obtained 
 by a mild tuning of the Yukawa coupling $Y_R \sim 0.1$.

\section{Gauge coupling unification}
In the orbifold GUT model, we assume that 
 a more fundamental extra-dimensional GUT theory 
 takes place at some high energy beyond the compactification scale.  
For the theoretical consistency of the model, 
 the gauge coupling unification should be realized 
 at some scale after taking into account 
 the contributions of Kaluza-Klein modes of the bulk
 gauge multiplet to the gauge coupling running \cite{GCU}.

In our setup, we take $v_{PS} < M_c$ and 
 the evolution of gauge couplings has three stages, 
 $G_{321}$ (SM+MSSM), $G_{422}$ (the PS stage) 
 and the PS stage with the Kaluza-Klein mode contributions. 
Since we have imposed the left-right symmetry, 
 the SU(2)$_L$ and SU(2)$_R$ gauge couplings 
 must coincide with each other at the scale $\mu=v_{PS}$. 
As a consequence, the PS scale is fixed 
 from the gauge coupling running in the MSSM stage.

In the $G_{321}$ stage, we have 
\bea
\frac{1}{\alpha_i (\mu)}=\frac{1}{\alpha_i(M)}
 +\frac{b_i}{2\pi} \mbox{ln}\left(
\frac{M}{\mu}\right); ~~(i=3,2.1), 
\eea
where $M=M_Z$, and $b_i$s are
\bea
b_3=-7,~b_2=-19/6,~b_1=41/10
\eea
for $M_Z \leq \mu \leq  M_{\rm SUSY}$. 
For $M_{\rm SUSY} \leq \mu \leq v_{PS}$, $M=M_{\rm SUSY}$ and
\bea
b_3=-3,~b_2=1,~b_1=33/5. 
\eea
At the PS scale, the matching condition holds 
\bea
\alpha_3^{-1}(v_{PS})&=&\alpha_4^{-1}(v_{PS})\nonumber\\
\alpha_2^{-1}(v_{PS})&=&\alpha_{2L}^{-1}(v_{PS})\nonumber\\
\alpha_1^{-1}(v_{PS})&=& 
 \frac{2\alpha_4^{-1}(v_{PS})+3 \alpha_{2R}^{-1}(v_{PS})}{5} 
\label{maching}
\eea
For $\mu \geq M_c$ in the PS stage, 
 the threshold corrections $\Delta_i$ due to Kaluza-Klein modes 
 in the bulk gauge multiplet are added, 
\bea
\frac{1}{\alpha_i (\mu)}=\frac{1}{\alpha_i(v_{PS})}
+\frac{b_i}{2\pi}\mbox{ln}\left(
\frac{v_{PS}}{\mu}\right)+\Delta_i.~~(i=4,2_L,2_R)
\eea
The beta functions from the matter and Higgs multiplets 
 on the PS brane are 
\be
 b_4=3, ~b_{2L}=b_{2R}=6 .
\label{lrs}
\ee
The Kaluza-Klein mode contributions are given by 
\bea
\Delta_i&=& \frac{1}{2\pi}b_i^{even}\sum_{n=0}^{N_l}
\theta(\mu-(2n+2)M_c)\mbox{ln}\frac{(2n+2)M_c}{\mu} \nonumber\\  
&+&
\frac{1}{2\pi}b_i^{odd}\sum_{n=0}^{N_l}
\theta(\mu-(2n+1)M_c)\mbox{ln}\frac{(2n+1)M_c}{\mu}
\eea
with 
\bea
b_i^{even}&=&(-8,-4,-4) , \nonumber\\ 
b_i^{odd}&=&(-8,-12,-12)  
\eea
under $G_{422}$. 

\begin{figure}[ht]
\begin{center}
{\includegraphics*[width=.6\linewidth]{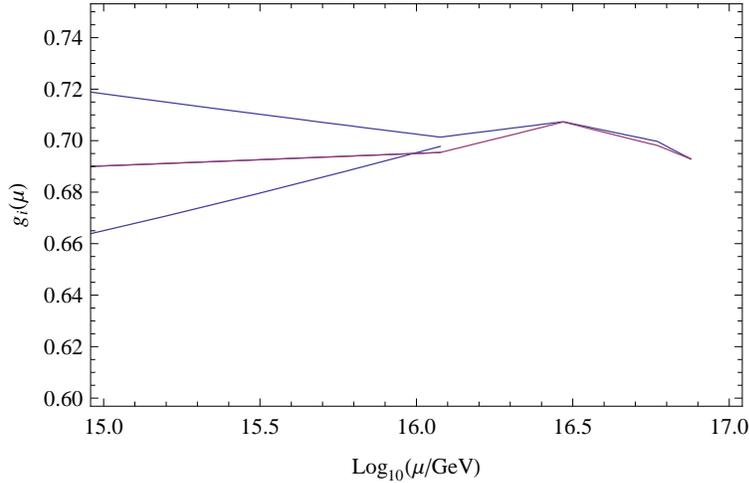}
\label{Fig}}
\caption{
Gauge coupling unification in the left-right symmetric case. 
Each line from top to bottom corresponds to 
 $g_3$, $g_2$ and $g_1$ for $ \mu < v_{PS}$,
 while  $g_3=g_4$ and $g_2=g_{2R}$ for $ \mu > v_{PS}$. 
Here, we have taken $M_c= 2.47 \times v_{PS}$.
}
\end{center}
\end{figure}

Fig.~1 shows the gauge coupling evolutions 
 for the left-right symmetric case. 
The PS symmetry breaking scale, $v_{PS}$, is determined 
 from the gauge coupling running in the MSSM stage 
 by imposing the matching condition, 
 $\alpha_2^{-1}(v_{PS})= \alpha_{2R}^{-1}(v_{PS}) 
 =(5\alpha_1^{-1}(v_{PS})-2\alpha_3^{-1}(v_{PS}))/3$, 
 and we find 
\bea 
 v_{PS}= 1.19 \times 10^{16} \; \mbox{GeV} , 
\eea 
 for the inputs, 
 $(\alpha_1(M_Z), \alpha_2(M_Z), \alpha_3(M_Z))
 = (0.01695, 0.03382, 0.1176)$ and $M_{\rm SUSY}=1$ TeV. 
For the scale $\mu \geq v_{PS}$, there are only two independent 
 gauge couplings $\alpha_4$ and $\alpha_2=\alpha_{2R}$, 
 so that the gauge coupling unification is easily realized 
 with a suitable $M_c$. 
In this figure, we have taken (corresponding to the result in the next section)
\bea 
 M_c = 2.47 \times v_{PS} = 2.95 \times 10^{16} \;  \mbox{GeV}  
\eea 
 and after including Kaluza-Klein threshold contributions 
 into the gauge coupling evolutions, 
 the gauge coupling unification is realized at 
\bea 
  M_{\rm GUT} = 7.54 \times 10^{16} \; \mbox{GeV} . 
\eea  
As $M_c$ is raised, $M_{\rm GUT}$ becomes smaller. 
As mentioned before, we assume that a more fundamental 
 SO(10) GUT theory takes place at $M_{\rm GUT}$.

\section{Gaugino mediation and sparticle masses} 
The origin of SUSY breaking and its mediation to the MSSM sector 
 is still a prime question in any phenomenological SUSY models. 
Since flavor-dependent soft SUSY breaking masses are 
 severely constrained by the current experiments, 
 a mechanism which naturally transmits SUSY breaking 
 in a flavor-blind way is the most favorable one.

In higher dimensional models, the sequestering \cite{RS-SUSY} 
 is the easiest way to suppress flavor-dependent SUSY breaking 
 effects to the MSSM matter sector. 
Since all matters reside on the PS brane in our model, 
 the sequestering scenario is automatically realized 
 when we simply assume a SUSY breaking sector 
 on the brane at $y=0$. 
The SO(10) gauge multiplet is in the bulk and 
 can directly communicate with the SUSY breaking sector  
 through the higher dimensional operator of the form,  
\bea 
 {\cal L} \sim \delta (y) 
 \int d^2 \theta \; 
  \frac{X}{M_5^2} 
 {\rm tr} \left[ {\cal W}^\alpha {\cal W}_\alpha \right],   
\eea 
where $X$ is a singlet chiral superfield 
 which breaks SUSY by its F-component VEV, $X= \theta^2 F_X$. 
Therefore, the bulk gaugino obtains the SUSY breaking soft mass, 
\bea 
 M_\lambda \sim \frac{F_X M_c}{M_5^2}
 \simeq \frac{F_X}{M_P} \left(\frac{M_5}{M_P} \right) ,  
\label{gaugino}
\eea
where $M_c$ comes from the wave function normalization 
 of the bulk gaugino, 
 and we have used the relation between the 4D and 5D Planck scales,  
 $M_5^3/M_c \simeq M_P^2$ ($M_P=2.4 \times 10^{18}$ GeV) 
 in the last equality. 
As usual, we take $M_\lambda =$100 GeV-1 TeV.  
Once the gaugino obtains non-zero mass, 
 SUSY breaking terms for sfermions are automatically 
 generated through the RGE from the compactification scale 
 to the electroweak scale. 
This scenario is nothing but the gaugino mediation \cite{gMSB} 
 and flavor-blind sfermion masses are generated 
 through the gauge interactions. 
In this setup, a typical gaugino mass in Eq.~(\ref{gaugino}) 
 is smaller than the gravitino mass $m_{3/2} \simeq F_X/M_P$ 
 by a factor $M_5/M_P < 1 $.

As discussed in Ref.~\cite{F-O1}, for $M_c=v_{PS}$, 
 we find that the right-handed slepton (normally, stau) 
 is the LSP, because the sfermion mass spectrum is 
 obtained from the boundary condition with vanishing soft masses 
 at $M_c=v_{PS}=1.19 \times 10^{16}$ GeV. 
This result is problematic for cosmology. 
As pointed out in Ref.~\cite{gMSB2}, 
 this stau LSP problem is cured by the soft mass RGE running 
 from the compactification scale to the GUT scale in a GUT model. 
In the following, we apply this idea to our model 
 with $M_c > v_{PS}$.

For the scale, $v_{PS} \leq \mu \leq M_c$,  
 we are in the PS stage and the RGEs of gaugino and sfermion masses 
 are given by 
\bea 
&&  \frac{d}{d t} \left( \frac{M_4}{\alpha_4} \right) = 
 \frac{d}{d t} \left( \frac{M_{2L}}{\alpha_{2L}} \right) = 
 \frac{d}{d t} \left( \frac{M_{2R}}{\alpha_{2R}} \right) =0,  \nonumber \\ 
&& \frac{d m^2_{\tilde{F}}}{d t} = 
 - \frac{15}{4 \pi} \alpha_4 M_4^2 
 - \frac{3}{2 \pi} \alpha_{2L} M_{2L}^2,  \nonumber \\ 
&& \frac{d m^2_{\tilde{F}^c}}{d t} =  
 - \frac{15}{4 \pi} \alpha_4 M_4^2 
 - \frac{3}{2 \pi} \alpha_{2R} M_{2R}^2 ,
\label{RGEs}
\eea
where $t=\ln(\mu/M_c)$, 
 $\alpha_4$ and $\alpha_{2L}=\alpha_{2R}$ are 
 the PS gauge coupling of the corresponding gauge groups 
 (whose RGE solutions are obtained in the previous section), 
 and $M_4$, $M_{2L}$ and $M_{2R}$ are 
 the corresponding gaugino masses. 
Sfermion mass spectrum is obtained by solving the RGEs 
 with the boundary conditions, 
 $m_{\tilde{F}}(M_c)=m_{\tilde{F}^c}(M_c)=0$. 
Analytic solutions of Eq.~(\ref{RGEs}) at $\mu =v_{PS}$ 
 are easily found: 
\bea 
  m^2_{\tilde{F}}(v_{PS}) &=& 
    \frac{5}{4} M_4^2(v_{PS}) \left[ 
     \left( \frac{\alpha_4(M_c)}{\alpha_4(v_{PS})} \right)^2 -1 \right] 
   +  \frac{1}{4} M_{2L}^2(v_{PS}) \left[ 
    \left( \frac{\alpha_{2L}(M_c)}{\alpha_{2L}(v_{PS})} \right)^2 -1 
 \right], \nonumber \\ 
  m^2_{\tilde{F}^c}(v_{PS}) &=& 
   \frac{5}{4} M_4^2(M_c) \left[ 
    \left( \frac{\alpha_4(v_{PS})}{\alpha_4(v_{PS})} \right)^2 -1 \right] 
   + \frac{1}{4} M_{2R}^2(v_{PS}) \left[ 
    \left( \frac{\alpha_{2R}(M_c)}{\alpha_{2R}(v_{PS})} \right)^2-1
  \right]. 
\eea 
Note that the PS model is unified into a more fundamental SO(10) model 
 and this unification leads to the well-known relation, 
\bea 
 \frac{M_4}{\alpha_4}  = 
 \frac{M_{2L}}{\alpha_{2L}}  = 
 \frac{M_{2R}}{\alpha_{2R}}  = 
 \frac{M_{1/2}}{\alpha_{\rm GUT}}  , 
\eea
where $M_{1/2}$ is the universal gaugino mass 
 at the unification scale. 
Thus, the formulas for sfermion masses are simplified as 
\bea 
 && m^2_{\tilde{F}}(v_{PS}) = m^2_{\tilde{F}^c}(v_{PS}) \nonumber \\
 &=&  
  \left( \frac{M_{1/2}}{\alpha_{\rm GUT}} \right)^2 
  \left[ \frac{5}{4} 
     \left( \alpha_4^2(M_c) - \alpha_4^2 (v_{PS}) \right)
  +  \frac{1}{4} 
     \left( \alpha^2_{2L}(M_c) - \alpha^2_{2L} (v_{PS}) \right) 
  \right]. 
\eea

\begin{figure}[ht]
\begin{center}
{\includegraphics*[width=.6\linewidth]{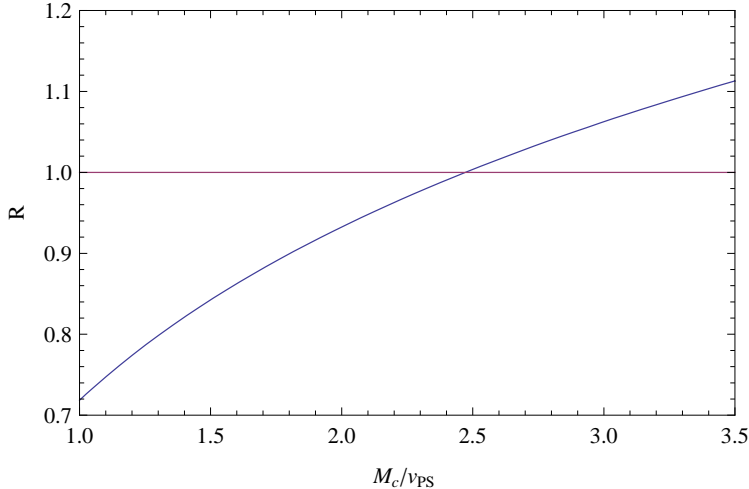}}
\caption{
The ratio, $R \equiv m_{\tilde{e^c}}^2/M_1^2$, 
 as a function of $M_c/v_{PS}$. 
Here, soft masses have been evaluated at $\mu=M_{\rm SUSY}$. 
$R=1$ when $M_c/v_{PS} = 2.47$. 
}
\end{center}
\end{figure}

Solving the RGEs in the MSSM with the universal boundary condition 
 $ m^2_{\tilde{F}}(v_{PS}) = m^2_{\tilde{F}^c}(v_{PS}) 
 = m_0^2$ at $\mu = v_{PS}$, we obtain the sfermion masses 
 at the electroweak scale. 
In our model, the PS scale is almost the same as the usual SUSY GUT 
 scale in the MSSM ($M_{\rm GUT} \simeq 2 \times 10^{16}$ GeV) 
 and the gauge couplings are roughly unified at the PS scale, 
 $ \alpha_4(v_{PS}) \simeq \alpha_{2}(v_{PS})= \alpha_{2R}$ 
 (see Fig.~1). 
Therefore, our study on the sfermion masses are 
 almost the same as the one usual in the constrained MSSM. 
For a small $\tan \beta$ (say, $\tan \beta=10$), 
 we neglect Yukawa coupling contributions to the soft masses 
 of right-handed sleptons, and the analytic solutions 
 of the MSSM RGEs are found to be 
\bea 
 M_1 (\mu) &=& \alpha_1(\mu) 
  \left(  \frac{M_{1/2}}{\alpha_{\rm GUT}} \right) , \nonumber \\ 
 m_{\tilde{e^c}}^2 (\mu) &=& 
  \left(  \frac{M_{1/2}}{\alpha_{\rm GUT}} \right)^2 
   \frac{2}{11} 
   \left[ \alpha_1^2(v_{PS})- \alpha_1^2(\mu) \right] 
   + m_0^2.  
\eea 
If $m_0$ is large enough, the slepton (stau) mass is bigger 
 than the bino mass ($M_1$). 
In our model, $m_0$ is given as a function of $M_c$. 
Fig.~2 shows the ratio, 
\bea 
  R \equiv \frac{m^2_{\tilde{e^c}}}{M_1^2},  
\eea
 as a function of $M_c/v_{PS}$. 
We can obtain $ R \geq 1$ for $M_c/v_{PS} \geq 2.47$.

Now, for $M_c/v_{PS} \geq 2.47$, the bino-like neutralino 
 will be the LSP and a good candidate for the cold dark matter 
 in cosmology \cite{DM}. 
For a small $ \tan \beta$, 
 the annihilation processes of the bino-like neutralino 
 are dominated by p-wave and are not so efficient. 
As a result, the neutralino relic density tends to exceed 
 the upper bound of the observed dark matter density. 
This problem can be avoided if the neutralino  is quasi-degenerate 
 with the next LSP slepton (stau), and the co-annihilation process 
 with the next LSP can lead to the right dark matter relic density. 
In our result, such a situation appears for 
 $M_c \simeq 2.47 \times v_{PS} \simeq 2.95 \times 10^{16}$ GeV. 
It would be interesting to note that 
 the discrepancy of the abundance of ${}^7$Li 
 between the observed values in WMAP and in metal poor halo stars 
 may be explained the degeneracy between 
 the LSP neutralino and stau \cite{Jittoh}.

\section{Conclusion}
We have considered SO(10) GUT in 5D, where the fifth dimension 
 is compactified on the $S^1/(Z_2\times Z_2^\prime)$ orbifold 
 with two inequivalent branes at the orbifold fixed points. 
All the matter and Higgs multiplets reside on the PS brane, 
 while the SUSY breaking sector is on the other brane. 
In this setup, we have two independent energy scales; 
 $v_{PS}$ at which the PS symmetry is broken on the PS brane
 and $M_c$, the inverse of radius of the fifth dimension. 
Requiring the left-right symmetry in the model, 
 the PS symmetry breaking scale is determined 
 from the MSSM gauge coupling runnings. 
For the case with $M_c > v_{PS}$, 
 we have investigated the gauge coupling unification 
 and the sparticle mass spectrum through the gaugino mediation. 
We have found that after incorporating threshold corrections 
 of Kaluza-Klein modes of the bulk gauge multiplets, 
 the gauge couplings can be successfully unified at $M_{\rm GUT}$, 
 whose scale is determined once $M_c$ is fixed. 
Also, we have found that an appropriate choice of $M_c$ 
 leads to the bino-like LSP neutralino quasi-degenerating with 
 the next LSP slepton (stau), so that the co-annihilation 
 between the LSP and next LSP provides the right relic abundance 
 of the neutralino dark matter.

\section*{Acknowledgments}
T.F. is grateful to S. Petcov for his hospitality at SISSA. 
N.O. would like to thank the Maryland Center for Fundamental Physics, 
and especially Rabindra N. Mohapatra for their hospitality 
and financial support during his stay. 
The works of T.F. and N.O. are supported in part by 
the Grant-in-Aid for Scientific Research from the Ministry 
of Education, Science and Culture of Japan 
(No. 20540282 and No. 18740170, respectively).


\end{document}